# *In situ* Monitoring of Composition and Sensitivity to Growth Parameters of Pulsed Laser Deposition

*Thomas Orvis[†], Harish Kumarasubramanian, Mythili Surendran[†], Shanmukh Kutagulla, Austin Cunniff, and Jayakanth Ravichandran[§*]*

Mork Family Department of Chemical Engineering and Materials Science, University of Southern California, 925 Bloom Walk, Los Angeles, CA 90089, USA

**ABSTRACT**

Complex oxide perovskites have been widely studied for their diverse functional properties. When dimensionally reduced to epitaxial thin films and heterostructures these properties are frequently tunable, and the symmetry-breaking inherent to thin film structures can result in the emergence of new, novel, phenomena and properties. However, the ability to control and harness these structures relies on an atomic-level understanding and control of the growth process, made challenging by the lack of suitable in situ compositional characterization tools. In this work, the compositional-dependence of $SrTiO_3$ on pulsed laser deposition growth parameters is investigated with in situ Auger electron spectroscopy and ex situ thin film x-ray diffraction, and verified with a simple escape depth model. We show that this is a suitable technique for monitoring subtle compositional shifts occurring during the deposition process, with broad implications for the continued development of thin film synthesis techniques.

**INTRODUCTION**

Epitaxial heterostructures of complex oxides are model systems for the realization of unconventional phases and the application of their functionalities to emerging technologies.[1-6] As these materials are frequently sensitive to subtle variations in structure and composition, their properties are highly dependent on the growth conditions, especially in high pressure techniques such as pulsed laser deposition (PLD).[7-9] The development of high pressure compatible *in situ* structural characterization techniques has allowed for greater understanding of how these growth conditions can be tailored to achieve atomic level control over thickness and structural phase stability.[10-12] However, a complete understanding of the inherent complexities of the growth process can only be achieved with similar purview and control over the composition.[13,14] To address this open issue requires the development of facile *in situ* elemental and chemical composition analysis tools operable during the deposition process. Advances in tools and techniques for *in situ* and real time elemental analysis will significantly enhance our ability to atomically engineer complex oxide heterostructures, thereby allowing the full realization of these materials' promise in computing, communication, sensing, and energy applications.

Pulsed laser deposition is a particularly popular approach for the synthesis of epitaxial complex oxides due to the simplicity of the method and the near-stoichiometric transfer of the components from the target-to-film without the need for significant optimization.[15,16] However, while stoichiometric transfer is possible, it has been repeatedly demonstrated that the actual composition of the film is often highly sensitive to the conditions used for its growth, including ambient pressure,[17,18] substrate temperature,[19] laser fluence,[20] and even laser spot size and plume position relative to the substrate.[21,22] While subtle shifts in composition may not be particularly important for the success of some thin films and heterostructures, there are many materials and heterostructures which have composition-dependent properties extremely sensitive to minor changes, such as the two-dimensionally conducting $SrTiO_3$-$LaAlO_3$ interface,[23,24] the relaxor ferroelectric $0.5Ba(Zr_{0.2}Ti_{0.8})O_3$–$0.5(Ba_{0.7}Ca_{0.3})TiO_3$,[25,26] and charge-ordered manganite $La_{0.4}Ca_{0.6}MnO_3$.[27] The inherent multidimensional-complexity of PLD kinetics makes the process of perfecting growth recipes laborious and time-consuming, as characterization of the films is most often *ex situ*, and even when conducted *in situ* typically requires interrupting the growth process, making real time process control an impossibility. Furthermore, the tangled interrelationships between different growth parameters can make optimizing the growth conditions for desired stoichiometry a long and arduous process. The net result is an unfortunately wide variability in results reported on complex oxide heterostructures, which suppresses the community's ability to fully understand the depth and breadth of the process-properties relationship.[28,29] Even though the scope of the current study is confined to PLD, we anticipate that the approaches laid out are broadly applicable to thin film growth methods.

There is a direct correlation between advances in *in situ* surface characterization tools and improvements in the quality and control of thin films and heterostructures. Our ability to observe and understand the structure of the films we are depositing relies on techniques that have been in use for over half a century, such as reflection high energy electron diffraction (RHEED) and low energy electron diffraction (LEED).[30] The development of these techniques, and their associated technologies, expanded their application to real time growth-rate observation in the early 1980's,[31] and the prohibitively high-pressure growth environment required for oxide thin film growth using PLD in the 1990's.[10,32] However, developing comparable tools for *in situ* compositional characterization of oxides has been more sedate, despite popular interest, because of the demanding design requirements for the harsh environment of oxide deposition systems.[33] Recently the approach to overcoming these constraints has been reenvisioned with the introduction of a differentially-pumped Auger electron spectroscopy (AES) probe for oxide deposition systems.[34] AES has been used as a powerful surface characterization technique for over half a century,[35] and using this probe it has recently been shown that *in situ* atomic-scale elemental composition during pulsed laser deposition is easily observed with this probe.[36] Here we demonstrate that this technique can be used for the real time observation of film composition during deposition. This significantly expands the narrow window used to understand thin film growth kinetics and the process-composition relationship, thereby providing unparalleled control over the quality and properties of complex oxide heterostructures synthesized using epitaxial growth techniques such as PLD.

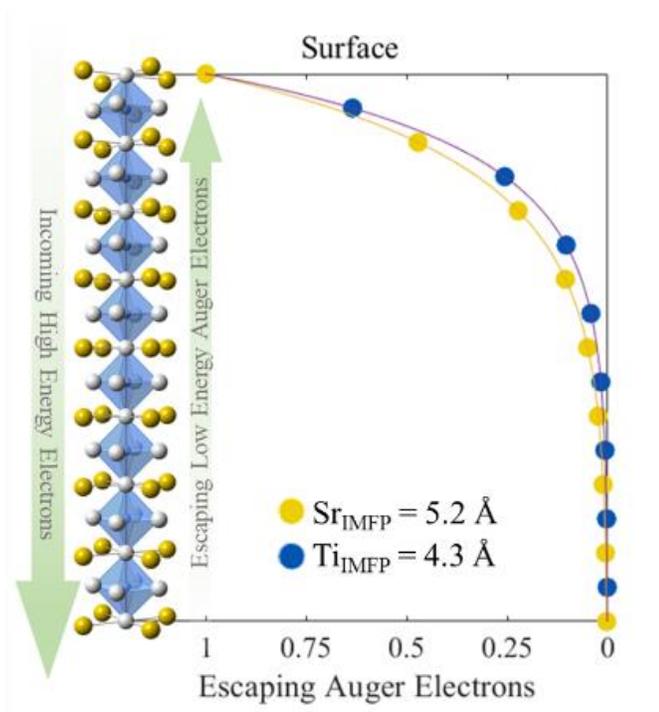

**Figure 1.** An illustration of Auger electrons' escape probability dependence on depth in the material and inelastic mean free path (IMFP), as well as relative position in an oriented layered structure such as SrO-terminated (100) $SrTiO_3$ as shown here. IMFPs are for the $Sr_{LMM}$ and $Ti_{MVV}$ transitions, calculated using Seah and Dench, 1979. Modeling the sensitivity of the probe for epitaxial perovskites is a simple function of layer depth and composition, as the number of escaping Auger electrons (signal intensity) will drop off exponentially at a rate determined by the electron's energy.

In this work, the composition of epitaxially-grown prototypical perovskite oxide $SrTiO_3$ (STO) is monitored *in situ* and in real time using AES during pulsed laser deposition as a function of growth parameters. Composition of the films is quantified using a combination of thin film x-ray diffractometry (XRD) and AES. Thin film XRD determines *c*-axis lattice expansion, which is known to shift as a function of composition.[37,38] AES composition quantification is verified using a parameter free escape depth model, the principle of which is illustrated in **Figure 1** and elucidated in the supporting information. A more complete understanding of the growth process is additionally developed by monitoring shifts in signal intensity with deposited thickness of the film. These results demonstrate our ability to actively monitor and control film composition during deposition, thereby allowing the direct study of PLD kinetics in a way up to now only possible piecemeal. The development of tools and techniques capable of monitoring the chemistry and composition of thin films on spatiotemporal scales pertinent to the process of their deposition provides us greater understanding, and therefore control, of the myriad complexities inherent to complex oxide thin film growth; the methods outlined herein allow this degree of observation, thereby providing an opportunity for the more complete understanding and application of thin film growth techniques.

**RESULTS & DISCUSSION**

We monitored the composition of 40 nm thick homoepitaxial STO thin films with *in situ* AES under a series of pressure and fluence conditions. We compared the composition of the single crystal (001) STO substrate before deposition, which was assumed to be stoichiometric, to that of the film after growth. The laser used a static spot size of 2.7 mm$^2$, and fluence adjustment was made solely by changing the energy of the laser output with the addition of fused quartz attenuators of various thicknesses in the optical path. The resulting fluences were 0.8, 1.8, and 2.8 J cm$^{-2}$, which are typical low, medium, and high fluences for our growth chamber, respectively. The growth pressures used were $10^{-4}$, $2 \times 10^{-3}$, and $5 \times 10^{-2}$ mbar, and were controlled by adjustment of O$_2$ flow rate and outlet area using a mass flow controller and butterfly valve, respectively.

Nine STO films were grown with the aforementioned growth parameters at 850°C with a target-substrate distance of 75 mm. RHEED oscillations are shown in the supplementary information. Thin film XRD was performed on the homoepitaxial (001) films to compare the *c*-axis lattice parameter to that of the substrate, with the expectation that off-stoichiometry films will exhibit composition-dependent *c*-axis expansion.[37,38] X-ray diffraction peaks are compared to AES E*N(E) spectra in **Figure 2**, with substrate and film peak positions marked. The obvious trend in film (002) peak position compared to the substrate shows that the films grown with a medium laser fluence of 1.8 J cm$^{-2}$ have the least c-axis lattice expansion, indicating that their compositions are closest to that of the substrate. The high laser fluence (2.8 J cm$^{-2}$) films have moderate expansion, while the low laser fluence (0.8 J cm$^{-2}$) films demonstrate the largest deviation in (002) peak position. The films grown with low laser fluence showed the largest variation in lattice expansion as a function of pressure. At high fluences, although the expansion was small, nominally the pressure dependence had an opposing trend to the low fluence series. The lattice parameter showed the least deviation from the bulk SrTiO$_3$ for films grown with medium fluence and low pressure.

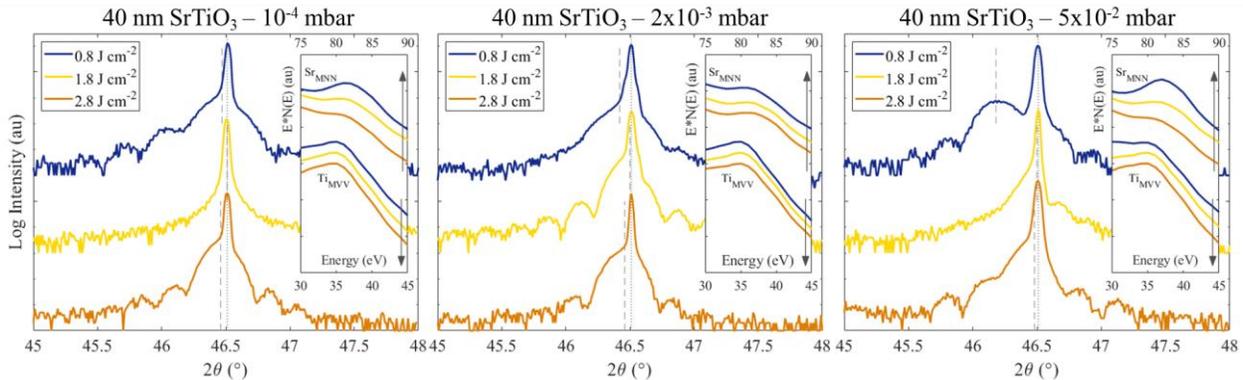

**Figure 2**: XRD and AES E*N(E) of (001) oriented homoepitaxial STO thin films. The primary substrate (002) peak at 2θ = 46.51° (lightly dashed line) is used as a reference to determine c-axis expansion of the thin films by comparing its position to the lower 2θ (002) peak position (heavily dashed line). A clear trend can be observed in which the medium laser fluence films (1.8 J cm$^{-2}$) have the least lattice expansion, while the low laser fluence films (0.8 J cm$^{-2}$) have the greatest, and higher pressure

exacerbates the low laser fluence trend. Comparing the Sr$_{MNN}$ and Ti$_{MVV}$ Auger spectra shows a similar visible trend, in which higher laser fluence increases the intensity of the Ti$_{MNN}$ peak, while lower laser fluence and higher pressure increase the intensity of the Sr$_{MNN}$ peak.

To further elucidate these observations, Auger spectra of the Sr$_{MNN}$ and Ti$_{MVV}$ peaks for each film are plotted as E*N(E) within insets beside the XRD peaks in Figure 2. The intensity of the Auger peaks correlate with composition, as peak signal is proportional to quantity of the species present. A simple, qualitative, inspection of the Auger peak intensities describes a relationship between growth parameters and composition in agreement with expectations. The intensity of the Sr$_{MNN}$ peak increases with decreasing laser fluence or increasing pressure. Likewise, the intensity of the Ti$_{MVV}$ peak increases with increasing the laser fluence. These trends indicate that the Sr/Ti ratio scales proportionately with pressure but inversely with fluence.

The dependence of film composition on laser fluence and growth pressure is well known.[39] Laser fluence is reported to change the stoichiometry of the plume largely in two ways: through preferential ablation and energy of the ablated species.[20,21] As STO has mixed ionic-covalent bonding states, with SrO exhibiting ionic bonding and TiO$_2$ covalent bonding, the energy-dependent ablation of these two species is not identical.[40] Specifically, Ti is more difficult to ablate, causing a lower laser fluence to result in a Sr-rich plume. The energy of the ablated species, and resulting plume shape, is more complex and ultimately dependent on numerous other factors such as spot shape and size, target-substrate distance, and pressure.[17,18,21,22,40] Pressure influences the propagation of the plume and diffusion of its species. Again, because of the discrepancy in atomic weight between Sr and Ti, the two species will behave differently within the plume. Lighter weight Ti will be more readily diverted due to collisions, and higher pressure will increase the probability of collision. Therefore, while higher pressure will increase collision frequency and thus diffusion for both species, their compositional profile within the plume will be different, and the stoichiometry of the plume-front which reaches the substrate will be more Sr-rich with high pressure. This principle is illustrated in **Figure 3** with a simple Monte Carlo simulation depicting the flight paths and final destinations of co-ablated Sr and Ti under two different pressure regimes. The difference in atomic mass between the two species results in diverging deposition patterns highly sensitive to background pressure. Details of the simulation can be found in the supporting information.

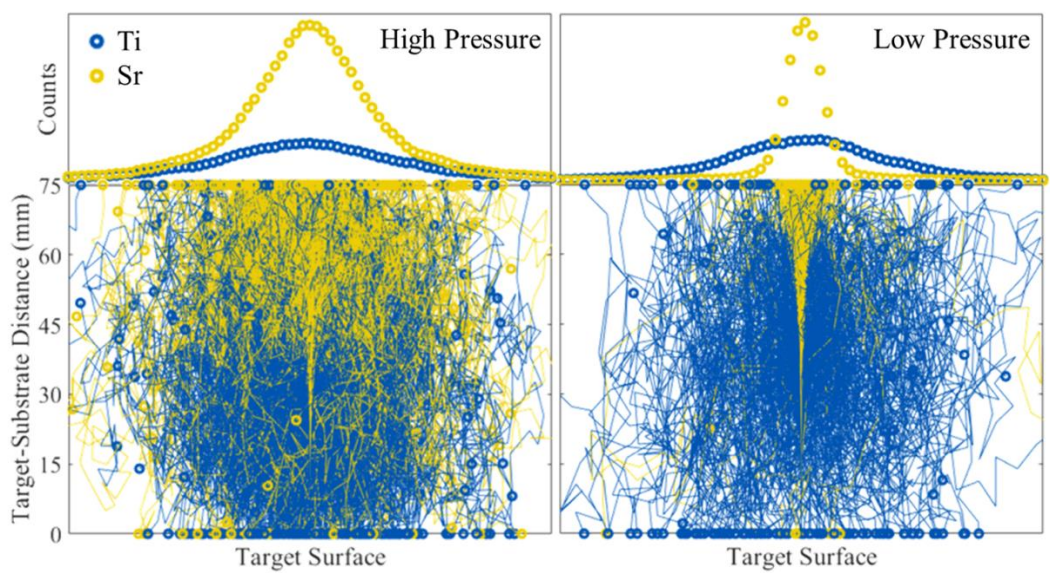

**Figure 3**: A simple Monte Carlo simulation of the pressure-dependent flight paths between target and substrate (bottom) and final particle distribution on the substrate surface (top) illustrates a significant deviation in deposition composition for Sr and Ti arising solely from the difference in atomic mass. Comparing high (left) versus low (right) pressure shows the dramatic influence of growth parameters on the resulting composition.

To better understand the observed trends in composition as a function of laser fluence and growth pressure, the collected AES results are quantified to qualitatively compare the composition information. Peak-to-peak (P2P) values, representative of the composition, were calculated from the summed Auger spectra for each set of growth parameters and compared with the *c*-axis lattice constant calculated from the XRD data, summarized in **Figure 4**. Demonstration of P2P calculations are illustrated in the supporting information. The correlation between Sr/Ti ratio and *c*-axis expansion for the lowest laser fluence series is clear, with all films showing a higher Sr/Ti ratio than the substrate and corresponding increase in lattice constant. The middle laser fluence films show a tendency toward Sr-richness with increasing pressure and an equivalently slight increase in lattice constant. The high laser fluence films show Ti-richness with higher pressure moving the Sr/Ti ratio towards stoichiometric, and the lattice constant likewise decreasing towards that of the substrate. Overall, the *in situ* AES signal indicates that film composition varies from ideal stoichiometry following the gross trends reported in the literature, corroborated by *c*-axis lattice expansion. They show that one could find an optimal set of fluence and pressure conditions to grow metal-stoichiometric $SrTiO_3$. It is worth noting that post annealing may be required for films grown at low pressures to achieve full oxygen stoichiometry. Presumably, the oxygen composition can be monitored using our technique, but is beyond the scope of the current study. These results demonstrate the viability of this probe for the *in situ* compositional analysis of thin films during deposition and opens the door for active process control and broader experimental inquiry.

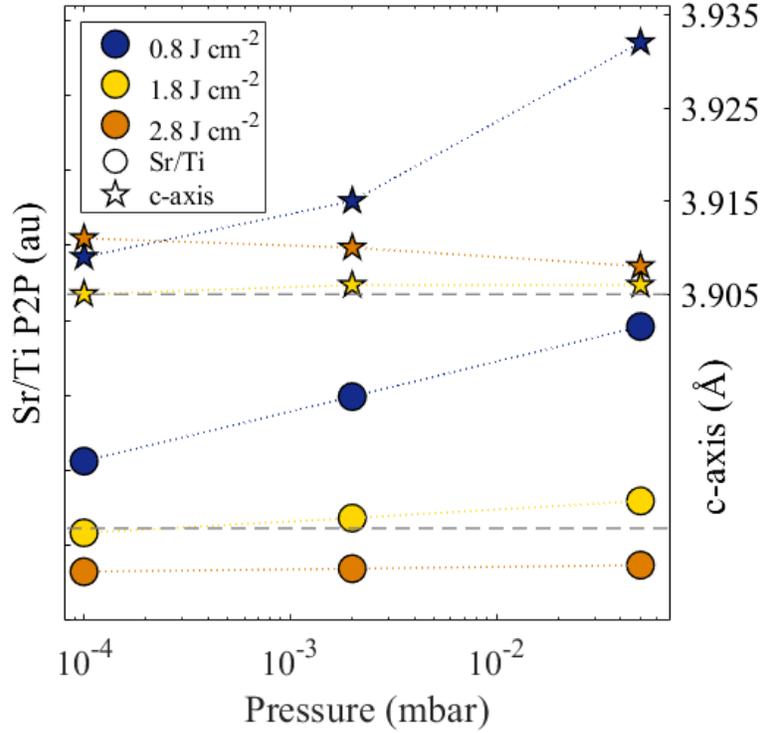

**Figure 4**: Auger signal intensity quantified as the ratio of the summed $Sr_{MNN}$ peak-to-peak value to that of $Ti_{MVV}$, for nine homoepitaxial STO thin films grown with the indicated laser fluences and pressures (bottom), as well as their c-axis lattice constants as derived from thin film XRD (top). Horizontal dashed lines indicate the Sr/Ti ratio and c-axis lattice constant of the single crystal STO substrate.

To develop a more comprehensive understanding of the compositional changes occurring during the deposition process, a homoepitaxial STO thin film was monitored with *in situ* AES while deliberately changing growth parameters to control composition. From the results of the pressure/fluence series presented in Figures 2 and 3, the initial growth parameters were chosen to maintain the same composition as the substrate in order to verify our ability to do so. STO was deposited two unit cells at a time, between which Auger spectra were collected for compositional analysis without altering any parameters of the growth. The deposition occurred with a substrate temperature of 850°C in $10^{-4}$ mbar flowing $O_2$. The Auger spectra were summed, and P2P values calculated, with Sr/Ti ratio as a function of overall film thickness and laser fluence presented in **Figure 5**. After the deposition of 20 unit cells of stoichiometric homoepitaxial STO using a laser fluence of 1.8 J cm$^{-2}$, fused quartz attenuators were placed in the laser path to reduce the laser fluence to 0.8 J cm$^{-2}$, making the deposition composition Sr-rich. The Sr/Ti ratio increased linearly until saturation after approximately 16 unit cells, and remained steady during the deposition of an additional 18 unit cells. The laser fluence was then returned to 1.8 J cm$^{-2}$ by removing the attenuators, and an additional 16 unit cells were deposited during which the Sr/Ti ratio steadily decreased, but did not return to the original ratio of the substrate. At this point, two unit cells of TiO$_2$ (equivalent to two half unit cells of STO) were deposited, bringing the Sr/Ti ratio below that of the substrate, indicating Ti-richness. An additional 28 unit cells of STO were then deposited with the same laser fluence of 1.8 J cm$^{-2}$, bringing the Sr/Ti ratio back to that of the substrate, demonstrating a return to the original composition after a total of 100 unit cells of STO.

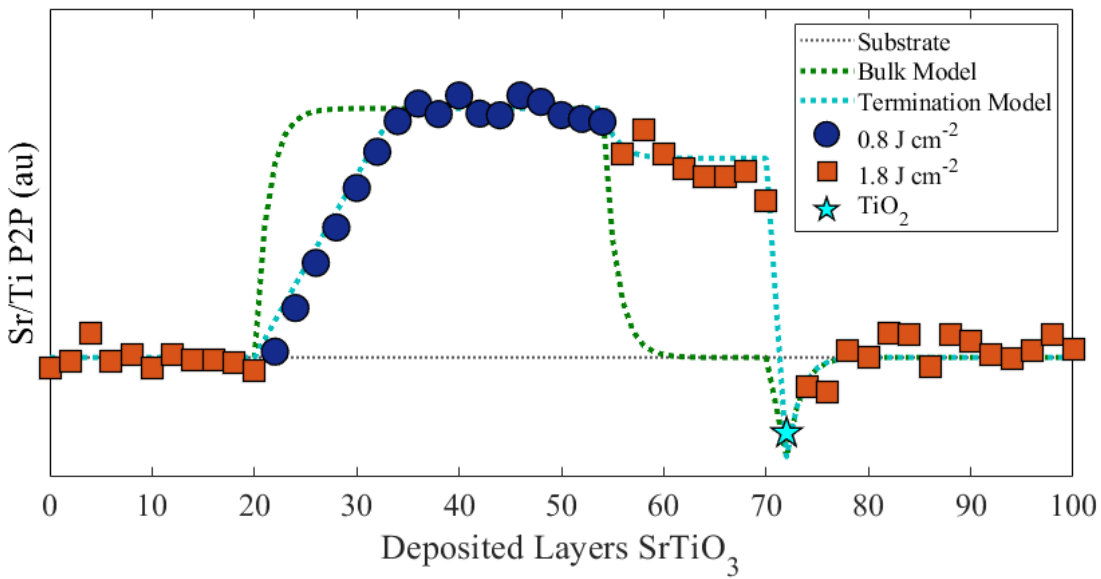

**Figure 5**: Sr/Ti ratio of the P2P calculated from summed Auger spectra collected after the deposition of every two unit cells of homoepitaxial STO, as a function of the number of deposited layers. The laser

fluence was changed periodically during the deposition, as indicated by the shape and color of the markers. A single deposition of two unit cells of $TiO_2$ (equivalent to two half unit cells of STO) was performed after depositing 70 unit cells of STO, demarcated as a star. The original composition of the STO substrate, as determined with AES, is indicated with a thin dashed gray line. The results of a parameter-free escape depth model used to explain the compositional variance as a function of deposition is shown as thick dashed lines, where the light line corresponds to a model which accounts for surface termination, while the dark line corresponds to a model which accounts only for the bulk.

If a homoepitaxial deposition has an off-stoichiometry plume, the resulting film must account for the excess species somehow. While some of the literature provides explanations for the manner in which the film incorporates an excess species,[37] many do not.[8,9,38] Regardless, the specifics will depend on the material and kinetics of the growth event in question. For STO, plausible scenarios to account for small amounts of excess Sr or Ti are the formation of interstitials and/or vacancies, or other types of point and line defects.[37] However, these fail to account for structure- and surface-dependent growth behavior such as termination, which are disproportionately influential in thin films.[12-14,41,42] Furthermore, the Auger electrons detected in AES typically have energies less than 2 keV, which corresponds to inelastic mean free paths (IMFPs) of a few nm at most.[36,43] This means that the extreme surface sensitivity of AES makes it particularly useful for observing these types of surface-dependent growth behaviors.

The trends observed in the Sr/Ti ratio of the growth shown in Figure 5 cannot be explained by bulk phenomena alone. If decreasing the laser fluence from 1.8 to 0.8 J cm$^{-2}$ changes only the bulk composition of the film the AES signal would saturate after approximately 5 unit cells, rather than the 15 unit cells of deposition that were required before signal saturation was observed. Likewise, when the laser fluence was returned to 1.8 J cm$^{-2}$ purely bulk phenomena would cause the Sr/Ti signal to return to that of the substrate after depositing only 5 or so unit cells of STO, rather than the lack of signal saturation observed after deposition of an additional 15 unit cells. These discrepancies are the result of the aforementioned small IMFP of low energy Auger electrons, as shown in Figure 1. For the $Sr_{MNN}$ and $Ti_{MVV}$ Auger electrons observed in this study, with energies of approximately 82 and 34 eV, and IMFPs of 5.2 Å and 4.3 Å,[43] respectively, 99% of the Auger signal comes from the top 5 unit cells of STO.[44] This means that the most reasonable explanation for a shift in signal which occurs over a duration greater than 5 unit cells is a shift in composition which takes place over the same period of time, and the simplest explanation to account for this sort of shift in composition is surface termination. Thinking of (001) oriented STO as alternating layers of SrO and $TiO_2$ means that the surface will consist of one, or a mixture of both, of these materials. Knowing that the initial surface termination of the substrate was $TiO_2$ due to the pre-growth chemical etching process,[45] a deposition of Sr-rich STO would gradually flip that termination to SrO as the surface would likely be the lowest energy location for the excess Sr. This process would cause the Sr/Ti signal to slowly increase over a much longer period than would be accounted for with a bulk model alone, and eventually plateau when the termination conversion was complete.

To quantitatively understand the observed differences in signal we applied a parameter-free escape depth model to compare the expected signal observed with and without surface termination being a factor to consider. Details of the model can be found in the supporting information. Comparison of the expected signal shifts with and without accounting for surface termination are shown in Figure 5. In the model which does not account for surface termination, the "Bulk Model," deposition of Sr-rich STO would increase the Sr/Ti ratio exponentially and then plateau, reaching saturation (99% of maximum signal) after the deposition of only five unit cells. Likewise, when stoichiometric STO was then deposited on top of the Sr-rich layers, the Sr-rich signal would again be masked after only five unit cells. If, at that point, an additional layer of TiO$_2$ was deposited the Sr/Ti signal would drop immediately and then recover as the double-layer of TiO$_2$ was buried. However, if we apply a model accounting for surface termination, the "Termination Model," deposition of Sr-rich STO will slowly increase the Sr/Ti signal as a function of excess Sr. After completing the SrO terminating layer, the signal would then increase further, dependent on the manner in which the excess Sr is incorporated into the film. If the composition of the film after switching to SrO termination is simply Sr-rich in the same proportion as that indicated by the rate of termination-switching, the signal would increase slightly then plateau. Returning to stoichiometric deposition, the Sr/Ti signal would then decrease as the excess Sr layers are buried, while maintaining the SrO termination. The deposition of TiO$_2$ is then used to force the termination back to that of the initial substrate. At this point, deposition of stoichiometric STO will bury the double TiO$_2$ layer while maintaining TiO$_2$ termination. During deposition on the original and final TiO$_2$ terminated surfaces the composition does not appear to change, which is in agreement with a stoichiometric deposition.

Comparison of the two models against the data in Figure 5 clearly shows that the termination model provides a better fit, with $\chi^2 = 0.011$, versus $\chi^2 = 0.370$ for the bulk model. Additionally, the rate of termination conversion observed with Sr-rich deposition indicates a Sr/Ti compositional ratio of ~1.06. Applying the same escape depth model to determine Sr/Ti signal ratios for thick film compositions finds agreement with the experiment shown in Figure 5. Applied to the Sr/Ti signal ratios of the films shown in Figure 2 and 4, a clear trend in composition and lattice parameter can be observed, presented in **Figure 6**. Additionally, the compositions and lattice parameters reported here are consistent with values found in the literature,[7,37] illustrated for comparison in Figure 6. These results indicate that the Auger probe used to collect *in situ* composition data in these experiments is not only capable of providing gross compositional information, but with appropriate calibrations also capable of quantifying *in situ* film composition.

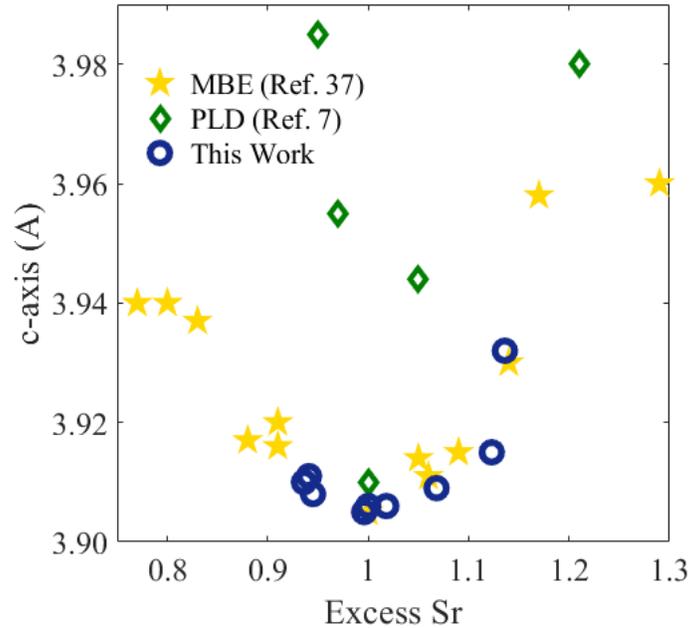

**Figure 6.** Normalized Sr/Ti Auger signal versus c-axis lattice constant for the homoepitaxial STO thin films presented in Figure 2 and 4, also plotted as excess Sr quantified from the parameter free escape depth model outlined in the text. For comparison, composition and *c*-axis data have been plotted from the literature,[7,37] showing agreement with our reported trends.

**CONCLUSION**

In summary, homoepitaxial thin films of STO were grown with PLD and characterized with *in situ* AES, and *ex situ* thin film XRD. Trends in the relationship between growth parameters and c-axis lattice parameter were explored, with expected correlations between the two found to agree with observed shifts in Auger spectra intensities. Further investigation revealed that the surface sensitivity of the Auger probe makes it suitable for quantification with a parameter free escape depth model, which accurately predicts signal as a function of changing growth parameters. Additionally, the model is used as the basis for compositional quantification of the grown films, exhibiting the suitability of this technique for *in situ* compositional analysis. The techniques outlined herein provide an additional axis of information to the increasingly complex field of epitaxial thin film growth, thereby expanding our capabilities for atomic-scale engineering and the pursuit of next-generation devices.

## ASSOCIATED CONTENT

**Supporting Information**. Escape Depth Model, RHEED Oscillations, Monte Carlo Plume Dispersion Simulation, Peak to Peak Calculations

## AUTHOR INFORMATION

**Corresponding Author**

* E-mail: j.ravichandran@usc.edu

**Present Addresses**

[†]Core Center for Excellence in Nano Imaging, University of Southern California, 925 Bloom Walk, Los Angeles, CA 90089, USA

[$]Ming Hsieh Department of Electrical and Computer Engineering, University of Southern California, 925 Bloom Walk, Los Angeles, CA 90089, USA

**Author Contributions**

The manuscript was written through contributions of all authors. All authors have given approval to the final version of the manuscript.

**Funding Sources**

This work was supported by the Air Force Office of Scientific Research and Army Research Office.

**ACKNOWLEDGMENT**

This work was supported by the Air Force Office of Scientific Research under contract FA9550-16-1-0335 and Army Research Office under Award No. W911NF-19-1-0137. T.O and M.S acknowledge the Andrew and Erna Viterbi Graduate Student Fellowship, and H. K. acknowledges the Annenberg Graduate Student Fellowship. The authors gratefully acknowledge support from Staib Instruments, as well as the benefit of discussions with Dr. Philippe G. Staib, Dr. Eric Dombrowski, and Laws Calley.